\documentstyle{aipproc}
\newcommand{\bq}{\begin{equation}}
\newcommand{\eq}{\end{equation}}

\newcommand\ds{\displaystyle}

\begin{document}
\begin{flushright}
{\tt hep-ph/9706205}
\end{flushright}

\title{Integral Relations for Twist 2 and Twist 3
Contributions  to Polarized Structure Functions\footnote{To appear in~:
Proc. of the 5th. Int. Conference on Deep Inelastic Scattering, Chicago,
April, 1997}}

\author{Johannes Bl\"umlein$^*$  and Nikolai Kochelev$^{*,\dagger}$}
\address{$^*$DESY -- Zeuthen, Platanenallee 6, D -- 15735 Zeuthen,
 Germany \\
$^{\dagger}$
Bogoliubov Laboratory of Theoretical Physics, JINR,
RU--141980 Dubna,  Russia }

\maketitle

\begin{abstract}
We discuss the relations between the twist 2 and twist 3 contributions to
polarized deep-inelastic scattering structure functions both for neutral
and charged current interactions which are predicted by the operator
product expansion in lowest order in QCD.
\end{abstract}

\section*{Introduction}
\noindent
In the case of polarized deep inelastic scattering the cross sections
depend on (up to) three unpolarized $F_i(x,Q^2)_{i=1}^3$ and five
polarized structure functions $g_j(x,Q^2)_{j = 1}^5$ in the limit of
vanishing fermion masses. The lowest twist contributions are those of
twist 2 for the structure functions $F_i(x,Q^2)$ and $g_{1,4,5}(x,Q^2)$.
The structure functions $g_2(x,Q^2)$ and $g_3(x,Q^2)$ contain as well
twist 3 terms~\cite{BK:AR,BK:BK2}.
In this note we give a summary on the relations between the twist 2
and twist 3 contributions to the different structure functions in
lowest order in QCD. We also comment on a sum rule which has been
derived in ref.~\cite{BK:ELT} recently.

\section*{Twist 2}
\noindent
For the twist 2 contributions to the structure functions
one may seek a partonic interpretation.
However, in lowest order in QCD only two
generic parton combinations exist to describe the polarized structure
functions~:
\begin{equation}
\label{E:BK:g15}
g_1(x,Q^2) \propto \Delta q(x,Q^2) + \Delta \overline{q}(x,Q^2)
~~~{\rm and}~~~
g_5(x,Q^2) \propto \Delta q(x,Q^2) - \Delta \overline{q}(x,Q^2).
\end{equation}
The remaining three structure functions are therefore related to
this basis by three linear operators. Two of the correponding relations
have been known for several years already,
the {\sc Dicus} relation~\cite{BK:DIC}
\begin{equation}
\label{E:BK:DIC}
g_4^i(x,Q^2)=2xg_5^i(x,Q^2),
\end{equation}
and the {\sc Wandzura-Wilczek} relation~\cite{BK:WW}
\begin{equation}
\label{E:BK:WW}
g_2^i(x,Q^2)=-g_1^i(x,Q^2)+\int_x^1\frac{dy}{y}g_1^i(y,Q^2).
\end{equation}
The third relation has been found only recently in ref.~\cite{BK:BK1}
\begin{equation}
\label{E:BK:BK1}
{g_3}^i(x,Q^2)=2x\int_x^1\frac{dy}{y^2}{g_4}^i(y,Q^2).
\end{equation}
Eqs.~(2--4) can  either be obtained  analyzing the polarized
structure functions by means of the operator product
expansion~\cite{BK:BK1,BK:BK2} or in applying the covariant parton
model~\cite{BK:CP}, cf.~\cite{BK:BK1}\footnote{A derivation of
eq.~(\ref{E:BK:WW}) using the latter method
was given in refs.~\cite{BK:ROB} before.}.
In the operator product expansion they result from equating different
expression for the matrix element $a_n$ of the symmetric part of the
quark operators in lowest order QCD,
see e.g.~\cite{BK:BK2}\footnote{The lowest moments of
$\left. g_i(x)\right|_{i=1}^5$ were studied in ref.~\cite{BK:FRA} and
agree with the
corresponding relations derived directly from eq.~(2--4).
}.
These relations between the different contributions to the longitudinal
and transverse spin projections of the hadronic tensor are illustrated
in Figure~1.
\renewcommand{\arraystretch}{2}
\[
\begin{array}{lclclc }

\vspace*{-5mm}
 & &                        & & & \hspace{-35mm}
{\sf Dicus}    \\

\vspace*{-5mm}
 & &    & &\hspace{15mm}   \swarrow  &
  \searrow  \\
     W^{\parallel}_{\mu\nu}
&=&
     i \varepsilon_{\mu\nu\alpha\beta} {\ds
\frac{q_{\alpha} P_{\beta}}{\nu} g_1(x)}
&+& {\ds \frac{P_{\mu} P_{\nu}}{\nu} g_4(x)}
&-      g_{\mu\nu} g_5(x)  \\
 & &\hspace{15mm}  \uparrow  & &\hspace{7mm}   \uparrow  &   \\
 & &{\sf  Wandzura-Wilczek} & & {\sf ref.~[6]}
&   \\
 & &\hspace{15mm}  \downarrow  & &\hspace{7mm}~\downarrow &   \\
{\ds W^{\perp}_{\mu\nu}} &=&{\ds
i \varepsilon_{\mu\nu\alpha\beta}
\frac{q_{\alpha}
S^{\perp}_{\beta}}{\nu} [g_1(x) + g_2(x)]} &+&
{\ds
\frac{P_{\mu} S^{\perp}_{\nu} + P_{\nu} S_{\mu}^{\perp}}{2 \nu}
g_3(x)}
&   \\
 & & \hspace{1cm}
{\ds \overbrace{\Delta q~~~~+~~~~\Delta \overline{q}}}
     &|& & \hspace*{-7mm}
 {\ds \overbrace{\Delta q~~~~-~~~~\Delta \overline{q}}}
\end{array}
\]

\vspace{2mm}
\noindent
\begin{center}
{\sf Figure~1~: Relations between the twist~2 contributions of the
polarized structure functions}
\end{center}

\vspace{2mm}
\noindent
Whereas the corresponding parts of
$W^{\parallel}_{\mu\nu}$ and ${\ds W^{\perp}_{\mu\nu}}$ are connected
by integral relations, those acting in either part are just
multiplications by a factor. For the valence parts this holds as well
for the two contributions to ${\ds W^{\perp}_{\mu\nu}}$.

\section*{Twist 3}
\noindent
For the twist 3 contributions to the structure functions $g_2(x,Q^2)$
and $g_3(x,Q^2)$, which emerge in the different neutral and
charged
current reactions, the operator product expansion~\cite{BK:BK2}
implies the relation~:
\begin{eqnarray}
\lefteqn{
\int_0^1dx x^n\{4g_5-\frac{n+1}{x}g_3\}^{\nu n-\nu p} =}
 \nonumber\\
& &~~~~~~~~~~~~~
\frac{12(n-1)}{n}
\int_0^1dx x^n\{ng_1+(n+1)g_2\}^{\gamma p-\gamma n},  n~=~2,4~...~.
\label{BK:MOM}
\end{eqnarray}
It results from equations between  differences of the matrix elements
$d_n$ of
the non-symmetric part of the quark operators in lowest order QCD.
Since one may express the twist 3 contributions to $g_2$ and $g_3$ by
\begin{eqnarray}
\label{E:BK:II}
g_2^{\rm III}(x,Q^2) &=& g_2(x,Q^2) + g_1(x,Q^2) - \int_x^1 \frac{dy}{y}
g_1(y,Q^2), \\
\label{E:BK:III}
g_3^{\rm III}(x,Q^2) &=& g_3(x,Q^2)  - 4 x \int_x^1 \frac{dy}{y}
g_5(y,Q^2),
\end{eqnarray}
the analytic continuation  of eqs.~(\ref{E:BK:II},\ref{E:BK:III})
in $n$ can be rewritten by
\begin{equation}
g_3^{{\rm III}, \nu n - \nu p}(x,Q^2) =
12 \Biggl [ x g_2^{\rm III}(x,Q^2)
 - \int_x^1 dy
g_2^{\rm III}(y, Q^2) \Biggr ]^{\gamma p - \gamma n}
\end{equation}
as a relation between twist 3 contributions {\it only}.

Recently, a sum rule for the valence part of the structure functions
$g_1(x,Q^2)$ and $g_2(x,Q^2)$
\begin{equation}
\int_0^1dx x(g_1^V(x)+2g_2^V(x))=0
\label{E:BK:etl}
\end{equation}
was discussed in ref.~\cite{BK:ELT}\footnote{
This sum rule was found  firstly in ref.~\cite{BK:A1C} for a
specific flavor combination.}.
We would like to investigate the relation of  eq.~(\ref{E:BK:etl})
to the operator product expansion.
Here one firstly meets the problem that  the valence parts
$g_1^V(x)$ and $g_2^V(x)$ cannot be isolated for electromagnetic
interactions from the complete structure functions and  a formulation
of eq.~(\ref{E:BK:etl}) with the help of the local operator product
expansion is thus not straightforward. On the other hand, one
may consider
\begin{eqnarray}
\lefteqn{
\int_0^1dxx^n(g_1^-(x,Q^2)+2g_2^-(x,Q^2)) =}
 \nonumber\\    & &~~~~~~~~~~~~
\sum_q \frac{((g_V^q)^2+(g_A^q)^2)
(nd_n^{-q}-(n-1)a_n^{-q})}{4(n+1)},~~~~n=1,3~...~,
\label{E:BK:etl1}
\end{eqnarray}
which results from eqs.~(61,62), ref.~\cite{BK:BK2}, in the
charged current case.
It is easily seen that the left-hand-side of eq.~(\ref{E:BK:etl1})
includes only valence  quark contributions and one  may even rewrite
eq.~(\ref{E:BK:etl1}) for individual quark flavors separately,
denoting the valence parts of the corresponding matrix elements by
$a_n^{Vq}$ and $d_n^{Vq}$, respectively.
For the first  moment one obtains
\begin{equation}
\int_0^1dxx(g_1^{Vq}(x,Q^2)+2g_2^{Vq}(x,Q^2))=\frac{e_q^2}{8}
d_1^{Vq}.
\label{E:BK:etl3}
\end{equation}
The right-hand-side of eq.~(\ref{E:BK:etl3})
 vanishes in the case of massless quarks,
because
\begin{eqnarray}
d_1^{Vq}(S^\beta P^\mu-  S^\mu P^\beta)
&=&  m_q \langle PS|\bar q i\gamma_5\sigma^{\beta\mu}q|PS \rangle
\nonumber\\
 &\equiv& \frac{m_q}{M}
(P^\beta S^\mu-P^\mu S^\beta)
\int_0^1dx(h_1^q(x)-\bar h_1^q(x))~.
\label{E:BK:em}
\end{eqnarray}
For the latter equation, see~\cite{BK:RLJ}.
$h_1^q(x)$, $\bar h_1^q(x)$ are the
quark and antiquark transversity functions,
respectively, which can be measured in the
Drell-Yan process.
The right-hand-side of eq.~(\ref{E:BK:em})
vanishes in the
limit $m_q \rightarrow 0$, which yields $d_1^{Vq} = 0$. Due to this,
eq.~(\ref{E:BK:em}), similar to the case of the Burkhardt--Cottingham
sum rule~\cite{BK:BC},
\begin{equation}
\int_0^1dxg^i_2(x,Q^2)=0,
\end{equation}
is not described by the operator
product expansion, but is formally consistent with it.

\end{document}